\documentclass[twocolumn,preprintnumbers,amsmath,amssymb]{revtex4}
\usepackage{amsmath}
\usepackage{amssymb}
\usepackage{mathrsfs}
\usepackage{graphicx}% Include figure files
\usepackage{dcolumn}% Align table columns on decimal point
\usepackage{bm}% bold math
\usepackage{color}
\usepackage{hhline}

\begin{document}
%\draft

\title{Spin-polarized scanning tunneling microscopy characteristics of skyrmionic spin structures exhibiting various topologies}

\author{Kriszti\'an Palot\'as$^{1,2}$}
\email{palotas@phy.bme.hu}
\author{Levente R\'ozsa$^{3}$}
\author{Eszter Simon$^{1}$}
\author{L\'aszl\'o Udvardi$^{1,4}$}
\author{L\'aszl\'o Szunyogh$^{1,4}$}

\affiliation{1 Budapest University of Technology and Economics, Department of Theoretical Physics,
Budafoki \'ut 8., H-1111 Budapest, Hungary\\
2 Slovak Academy of Sciences, Institute of Physics, Department of Complex Physical Systems,
Center for Computational Materials Science, SK-84511 Bratislava, Slovakia\\
3 Hungarian Academy of Sciences, Wigner Research Center for Physics,
Institute for Solid State Physics and Optics, P.O. Box 49, H-1525 Budapest, Hungary\\
4 Budapest University of Technology and Economics, MTA-BME Condensed Matter Research Group,
Budafoki \'ut 8., H-1111 Budapest, Hungary}

\date{\today}

\begin{abstract}

The correct identification of topological magnetic objects in experiments is an important issue. In the present work we
report on the characterization of metastable skyrmionic spin structures with various topological charges ($Q=-3,-2,-1,0,+1,+2$)
in the (Pt$_{1-x}$Ir$_{x}$)Fe/Pd(111) ultrathin magnetic film by performing spin-polarized scanning tunneling microscopy
(SP-STM) calculations. We find that an out-of-plane magnetized tip already results in distinguished SP-STM contrasts for the
different skyrmionic structures corresponding to their symmetries. Our work also establishes an understanding of the
relationship between in-plane SP-STM contrasts and skyrmionic topologies through an investigation of the variation of the in-plane
angle between the spins along the perimeter of the structures, which can be characterized by the local vorticity or linear density
of the winding number. For spin structures exhibiting a uniform sign of the local vorticity throughout the whole skyrmionic area,
we demonstrate that (i) $|Q|$ can be determined from a single SP-STM image taken by any in-plane magnetized tip,
and (ii) an in-plane tip magnetization rotation provides the sign of $Q$ independently of the sign of the effective spin
polarization in the tunnel junction. We also discuss cases where the local vorticity is changing sign.
Finally, by increasing the Ir content of the PtIr overlayer, we find an appearing secondary outer ring in-plane SP-STM contrast
that is indicative of attractive skyrmions or antiskyrmions.

\end{abstract}

\maketitle

\section{Introduction}
\label{sec_int}

Magnetic skyrmions correspond to specific spin configurations in magnetic materials characterized by a finite topological charge
\cite{nagaosa13}. Skyrmions may order into a hexagonal lattice and represent a stable thermodynamic phase \cite{bogdanov94},
the presence of which has been experimentally observed in a wide array of materials in bulk or thin film form
\cite{muhlbauer09,munzer10,seki12,adams12,yu10,yu11,huang12,kezsmarki15,tokunaga15}. Skyrmions may also appear as localized
metastable states in the collinear phase of magnets \cite{leonov16}. This property turns them into ideal candidates as bits of
information in future technological applications \cite{fert13,zhang15}. The experimental observation of such isolated skyrmions
is primarily connected to ultrathin film systems \cite{moreau16,romming13,hsu16elec,woo16}. The understanding of skyrmion
formation in these materials is also supported by computational efforts ranging from \textit{ab initio} calculations determining
the interaction parameters \cite{heinze11,polesya14,dupe14,simon14,polesya16} to studies focusing on finite-temperature effects
regarding the skyrmion stability \cite{hagemeister15,rozsa-sk1,rohart16,lobanov16}.

As real-space spin structures, isolated skyrmions can conveniently be imaged by using spin-polarized scanning tunneling microscopy
(SP-STM) \cite{bergmann14,romming15prl}. The controlled manipulation (creation and annihilation) of isolated skyrmions using an
SP-STM tip has been demonstrated in Refs.\ \cite{romming13,hsu16elec}, which may be the key for writing and deleting information
in magnetic media in future applications.

Most experimental observations of skyrmions in ultrathin films so far are connected to systems with strong
Dzyaloshinsky-Moriya interactions \cite{Dzyaloshinsky,Moriya}, where all skyrmions possess the same topological charge
\cite{leonov16}. In comparison, it was recently demonstrated that frustrated Heisenberg exchange interactions may lead to the
stabilization of localized spin configurations with different topological charges \cite{Okubo,Leonov,Lin}. Investigating the
SP-STM images of such structures is also worthwhile to consider. Based on ab initio
calculations performed for a Pd/Fe bilayer on Ir(111) surface, Dup\'{e} et al.\ \cite{dupe16stm} reported SP-STM contrast
characteristics for a set of metastable skyrmionic structures, where they identified circular contrasts using an out-of-plane
magnetized tip and two types of contrasts employing an in-plane magnetized tip: (i) a two-lobes contrast for spin structures with
topological charge $|Q| = 1$ and (ii) a four-lobes contrast for a higher-order antiskyrmion with $|Q| = 2$.
In Ref.\ \cite{rozsa-sk3}, other types of metastable skyrmionic structures were also investigated in the
(Pt$_{1-x}$Ir$_{x}$)Fe/Pd(111) ultrathin magnetic film, and it was discussed how their shapes become distorted due to the
interplay between the Dzyaloshinsky-Moriya and the frustrated Heisenberg exchange interactions. These observations enable
the generalization of the findings of Dup\'{e} et al.\ \cite{dupe16stm} for the SP-STM contrasts of topologically distinct
skyrmionic structures.

By employing SP-STM calculations on the metastable skyrmionic spin structures with various topological charges
($Q=-3,-2,-1,0,+1,+2$) taken from Ref.\ \cite{rozsa-sk3}, our work establishes a connection between SP-STM contrasts and
skyrmionic topologies, most importantly through an investigation of the variation of the in-plane angle between the spins
along the perimeter of the structures, which can be characterized by the local vorticity or linear density of the winding number.
Our findings are expected to be applicable to the topological characterization of skyrmionic spin structures based on
experimentally measured SP-STM images using a series of in-plane tip magnetization orientations. In ideal cases characterized by a
uniform sign of the local vorticity throughout the whole skyrmionic area, we propose that the magnitude and the sign of $Q$ of the
topological object can be determined. A measured series of in-plane SP-STM contrasts are also expected to help in identifying spin
structures if the local vorticity is not uniform. Such a scenario is expected in anisotropic environments of skyrmion formation,
e.g., in reconstructed or confined film geometries \cite{hsu16elec,hagemeister16anis}, where arbitrary skyrmionic shapes with
complex domain wall structures can be found.

Besides stabilizing skyrmionic structures with different topological charges, frustrated Heisenberg exchange interactions also
modify the shape of isolated skyrmions with $Q=-1$. The characteristic feature is an oscillation of the spin around the direction
of the background magnetization \cite{Leonov}, which shows up as a sequence of sign changes in the in-plane spin component
\cite{rozsa-sk2}. This oscillation indicates a short-range attractive interaction between skyrmions or antiskyrmions.
In our SP-STM calculations, the sign changes of the in-plane spin component appear as a secondary outer ring in the contrast.
This contrast feature can be tuned by the Ir content of the PtIr overlayer, which affects the exchange interactions in the Fe
layer \cite{rozsa-sk2}.

The paper is organized as follows. In the Theory section the employed SP-STM calculation method and tunneling
parameters are described, and the topological charge and the vorticity of skyrmionic structures in magnetic films are defined,
providing also a theoretical connection between SP-STM contrasts and the local vorticity.
After that, calculated SP-STM images are presented and discussed, and conclusions are drawn on the relationship between
the SP-STM contrast and the skyrmionic topology as well as on the SP-STM detection of attractive skyrmionic objects.

\section{Theory}
\label{sec_met}

\subsection{SP-STM}
\label{sec_met_stm}

For the calculation of the SP-STM images of the skyrmionic spin structures, the three-dimensional Wentzel-Kramers-Brillouin
(3D-WKB) approximation \cite{palotas14fop} of electron tunneling has been employed, where the tunneling current at the tip apex
position $\mathbf{R}_{TIP}$ and bias voltage $V$ is calculated as the superposition of 1D-WKB contributions from the sample
surface atoms (sum over $a$) as \cite{palotas11stm}
\begin{eqnarray}
&&I\left(\mathbf{R}_{TIP},V\right)=\epsilon^2\frac{e^2}{h}\sum_a\int_0^V dU\nonumber\\
&\times&\exp\left[-\sqrt{\frac{8m}{\hbar^2}\left(\frac{\Phi_S+\Phi_T+eV}{2}-eU\right)}|\mathbf{R}_{TIP}-\mathbf{R}_a|\right]\nonumber\\
&\times&n_S^a\left(E_F^S+eU\right)n_T\left(E_F^T+eU-eV\right)\nonumber\\
&\times&\left[1+P_S^a\left(E_F^S+eU\right)P_T\left(E_F^T+eU-eV\right)\cos\phi_a\right].
\label{Eq_current}
\end{eqnarray}
Here, the exponential factor describes the tunneling transmission, where all electron states are assumed as exponentially
decaying spherical states \cite{tersoff83,tersoff85,heinze06} with an effective rectangular potential barrier in the vacuum
between the sample and the tip. The electronic structures enter the model by considering $n_{S(T)}$ the atom-projected charge
density of states and $P_{S(T)}$ the spin polarization of the sample surface (S) and the tip apex (T). $\phi_a$ is the angle of
the localized magnetic moment of surface atom $a$ with respect to the tip magnetization direction, $e$ is the elementary charge,
$h(\hbar)$ is the (reduced) Planck constant, $m$ is the electron's mass, and $\Phi_{S(T)}$ and $E_F^{S(T)}$ denote the electron
work function and the Fermi energy of the sample surface (tip), respectively. The $\epsilon^{2}e^{2}/h$ factor ensures the
correct dimension of the current. The value of $\epsilon$ has to be determined by comparing the calculated results of the
charge current with experiments, or with calculations using standard methods, e.g., the Bardeen approach \cite{bardeen61}. In our
calculations $\epsilon=1$ eV has been chosen \cite{palotas12orb} that gives comparable current values with those obtained by
the Bardeen method implemented in the BSKAN code \cite{hofer03pssci,palotas05}. Note that the choice of $\epsilon$ has no
qualitative influence on the reported SP-STM contrasts and conclusions.

In the present work, SP-STM images correspond to constant-current surfaces calculated at the bias voltage $V=0$ V, where
Eq.\ (\ref{Eq_current}) takes the form \cite{heinze06}:
\begin{eqnarray}
I\left(\mathbf{R}_{TIP}\right)&\propto&\sum_a\exp\left[-\sqrt{8m\Phi/\hbar^2}|\mathbf{R}_{TIP}-\mathbf{R}_a|\right]\nonumber\\
&\times&\left[1+P_S P_T\cos\phi_a\right],
\label{Eq_current1}
\end{eqnarray}
assuming $P_S^a=P_S$ for all surface atoms and $\Phi=\Phi_S=\Phi_T$. Motivated by a recent work \cite{dupe16stm}, we choose the
effective spin polarization of $P_{eff}=P_S P_T=\pm 0.4$ and consider the effect of its sign on the SP-STM contrasts.
Electron work functions of $\Phi_S=\Phi_T=5$ eV were taken. With the selected parameters, the current value $I=10^{-4}$ nA of the
constant-current surfaces corresponds to about 6 \AA\;minimal tip-sample distance and corrugation values between
30 and 40 pm.
Note that smaller corrugation values found in an experiment \cite{romming15prl} are either due to a different $P_{eff}$ magnitude
or to a larger tip-sample separation in the experiment, for a theoretical explanation of the latter effect see, e.g.,
Ref.\ \cite{palotas13contrast}. $P_{eff}$ also plays a crucial role in spin-polarized scanning tunneling spectroscopy
\cite{palotas12sts}.

\subsection{Skyrmionic topology}
\label{sec_met_top}

For the characterization of the observed isolated skyrmionic objects, we rely on the topological charge $Q$, which expresses
how many times the spin vectors span the whole unit sphere. $Q$ is defined as
\begin{equation}
Q=\frac{1}{4\pi}\int\mathbf{S}\cdot\left(\partial_{x}\mathbf{S}\times\partial_{y}\mathbf{S}\right)\textrm{d}x\textrm{d}y,
\label{Eq_Q}
\end{equation}
where $\mathbf{S}=[\sin\Theta\cos\Phi,\sin\Theta\sin\Phi,\cos\Theta]$ is the unit vector of local magnetization and the
integral has to be performed over the area of the localized spin structure in the surface ($xy$) plane. The integral can be
transformed to a form considering surface polar coordinates $(r,\varphi)$ \cite{nagaosa13,rozsa-sk1},
\begin{equation}
Q=\frac{1}{4\pi}\int_{0}^{\infty}\int_{0}^{2\pi}(\partial_{r}\Theta\partial_{\varphi}\Phi-\partial_{\varphi}\Theta\partial_{r}\Phi)\sin\Theta\textrm{d}\varphi\textrm{d}r.
\label{Eq_Qpolar}
\end{equation}
It was demonstrated in Ref.\ \cite{rozsa-sk3} that due to the presence of the Dzyaloshinsky-Moriya interaction in the system,
not all of the observed skyrmionic objects possess a circular shape. The integral over $\varphi$ in Eq.\ (\ref{Eq_Qpolar}) may be
performed along the contour lines $r(\varphi)$ of $\Theta$, defined as being perpendicular to the gradient at all points,
\begin{equation}
\frac{\textrm{d}r(\varphi)}{\textrm{d}\varphi}=-\frac{\partial_{\varphi}\Theta}{\partial_{r}\Theta}.
\label{Eq_rphi}
\end{equation}
Here we only consider single-domain skyrmionic structures, where $\partial_{r}\Theta$ remains nonzero in the whole
considered configuration, and all the contours still only wind once around the origin. By introducing the local vorticity along
the contour lines,
\begin{equation}
\mathscr{M}(r,\varphi)=\frac{\textrm{d}\Phi(r,\varphi)}{\textrm{d}\varphi}=\partial_{\varphi}\Phi+\partial_{r}\Phi\frac{\textrm{d}r(\varphi)}{\textrm{d}\varphi},
\label{Eq_m}
\end{equation}
and substituting into Eq.\ (\ref{Eq_Qpolar}) one arrives at the expression
\begin{eqnarray}
Q&=&\frac{1}{2}\int_{0}^{\infty}\left[\frac{1}{2\pi}\int_{0}^{2\pi}\mathscr{M}(r,\varphi)\textrm{d}\varphi\right]\sin\Theta\partial_{r}\Theta\textrm{d}r\nonumber
\\
&=&-\frac{1}{2}[\cos\Theta(r)]_{0}^{\infty}m,
\label{Eq_Q2}
\end{eqnarray}
where the vorticity is defined as
\begin{equation}
m=\frac{1}{2\pi}\int_{0}^{2\pi}\mathscr{M}(r,\varphi)\textrm{d}\varphi.
\label{Eq_m2}
\end{equation}
The integral on the right-hand side of Eq.\ (\ref{Eq_m2}) gives an integer, expressing how many times and in which direction
the in-plane component of the spins rotates around the circle. Since there are no topological defects in the system, the value of
$m$ does not depend on the choice of the contour line, and is actually the same when performing the integral along an arbitrary
closed curve in the surface plane enclosing the center of the localized spin configuration.

Finally, the relationship between the local vorticity $\mathscr{M}$ and the topological charge density $\mathscr{Q}$ is
\begin{equation}
\mathscr{Q}(r,\varphi)=\frac{1}{4\pi r}\partial_{r}\Theta\sin\Theta\mathscr{M}(r,\varphi).
\label{Eq_Qdens}
\end{equation}
Since $\partial_{r}\Theta$ is negative for the skyrmionic spin structures considered in this paper, i.e., $\Theta(r=0)=\pi$ and
$\Theta(r=\infty)=0$, the signs of the topological charge density and the local vorticity are the opposite,
\begin{equation}
\mathrm{sgn}\mathscr{Q}(r,\varphi)=-\mathrm{sgn}\mathscr{M}(r,\varphi),
\label{Eq_m3}
\end{equation}
and, correspondingly, the relation between the topological charge and the vorticity is $Q=-m$ \cite{rozsa-sk3}.

\subsection{Relation between SP-STM and skyrmionic topology}
\label{sec_met_stmtop}

In SP-STM with an in-plane magnetized tip of the orientation $\mathbf{e}_{TIP}=(\cos\varphi_{TIP},\sin\varphi_{TIP},0)$,
the tunneling current is proportional to (cf.\ Eq.\ (\ref{Eq_current1}))
\begin{eqnarray}
I(\mathbf{R}_{TIP})\propto\mathbf{S}\cdot\mathbf{e}_{TIP}=\sin\Theta\cos(\Phi-\varphi_{TIP}).
\label{Eq_curtop}
\end{eqnarray}
Suppose that we select a single point on the constant-current surface (SP-STM image) with a current value of
$I(\mathbf{R}_{TIP})$, and by infinitesimally rotating the tip magnetization by $\textrm{d}\varphi_{TIP}$, we follow the
trajectory of the point on the constant-current surface along the $r(\varphi)$ contour of $\Theta$.
This procedure can mathematically be expressed as
\begin{eqnarray}
\textrm{d}I(\mathbf{R}_{TIP})&\propto&\mathscr{M}(r,\varphi)\textrm{d}\varphi-\textrm{d}\varphi_{TIP}=0,\nonumber
\\
\frac{\textrm{d}\varphi}{\textrm{d}\varphi_{TIP}}&=&\mathscr{M}^{-1}(r,\varphi).\label{Eq_v}
\end{eqnarray}
Equation (\ref{Eq_v}) means that the angular velocity $\textrm{d}\varphi/\textrm{d}\varphi_{TIP}$ of contrast features in
constant-current images obtained by in-plane tip magnetization rotation equals to the inverse of the local vorticity.
This enables the direct extraction of information in SP-STM experiments on the local vorticity $\mathscr{M}(r,\varphi)$ or on the
topological charge density $\mathscr{Q}(r,\varphi)$ using the proportionality relation in Eq.\ (\ref{Eq_Qdens}).

\section{Results and discussion}
\label{sec_res}

\begin{figure}[t]
\includegraphics[width=1.00\columnwidth,angle=0]{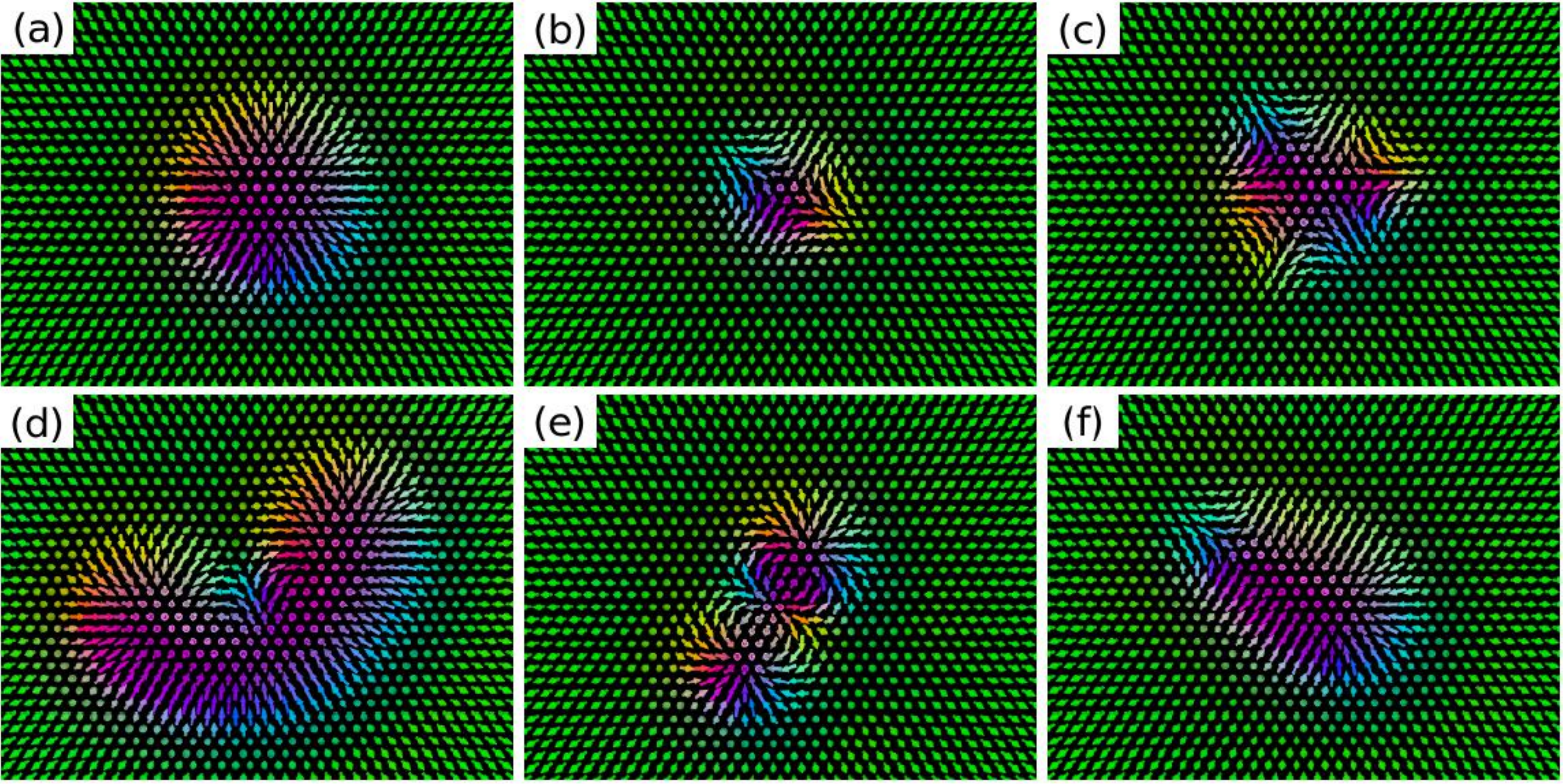}
\caption{\label{Fig1} (Color online) Metastable localized skyrmionic spin configurations with different topological charges in
the (Pt$_{0.95}$Ir$_{0.05}$)Fe/Pd(111) ultrathin magnetic film \cite{rozsa-sk3}:
(a) skyrmion with $Q=-1$, (b) antiskyrmion with $Q=1$, (c) antiskyrmion with $Q=2$,
(d) skyrmion with $Q=-2$, (e) skyrmion with $Q=-3$, (f) ``chimera'' skyrmion with $Q=0$.
The value of the external field is $B=0.23$ T (a)-(d),(f) and $B=2.35$ T (e); the ground state is field-polarized for $B>0.21$ T.}
\end{figure}

\begin{figure}[t]
\includegraphics[width=1.00\columnwidth,angle=0]{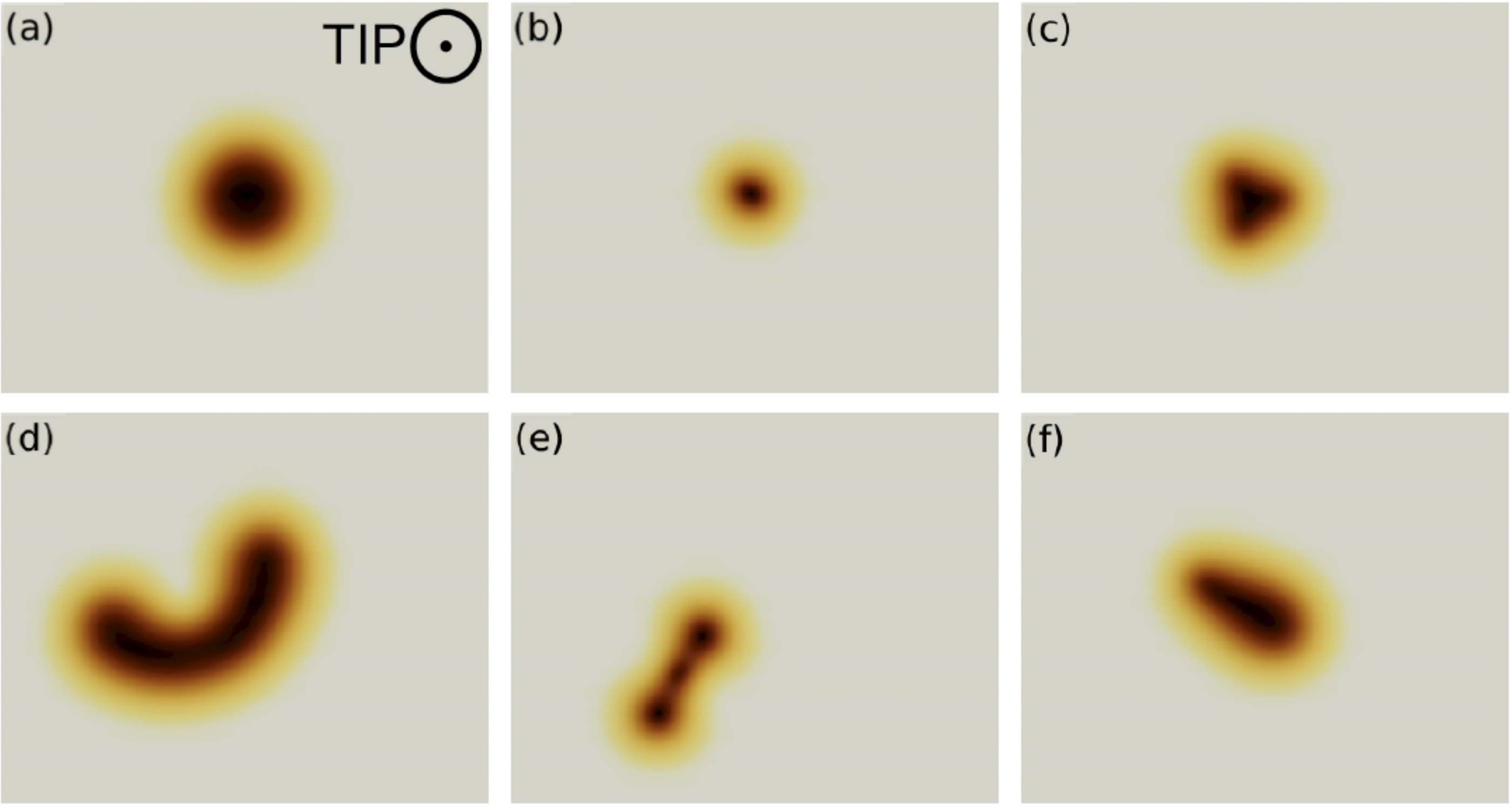}
\caption{\label{Fig2} (Color online) Calculated SP-STM images of the set of skyrmionic spin configurations shown
in Fig.\ \ref{Fig1} using an out-of-plane magnetized tip (pointing to the $+z$ $[111]$ direction as illustrated in (a))
with $P_{eff}=+0.4$. The color scale and image areas are the same for all skyrmionic structures.}
\end{figure}

\begin{figure}[t]
\includegraphics[width=1.00\columnwidth,angle=0]{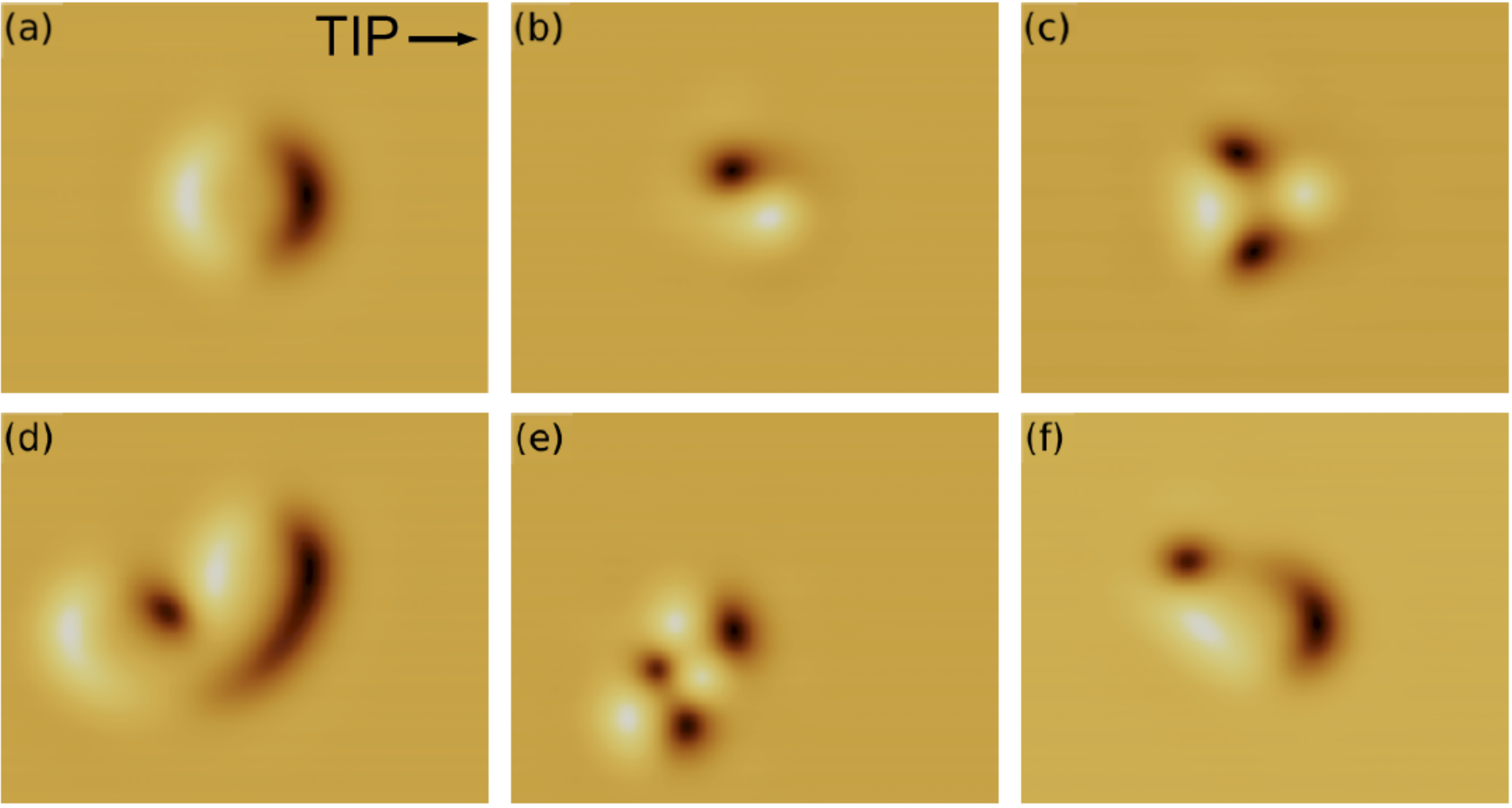}
\caption{\label{Fig3} (Color online) Same as Fig.\ \ref{Fig2} using an in-plane magnetized tip (pointing to the $+x$
$[1\bar{1}0]$ direction as illustrated in (a)) with $P_{eff}=+0.4$.}
\end{figure}

\begin{figure}[t]
\includegraphics[width=1.00\columnwidth,angle=0]{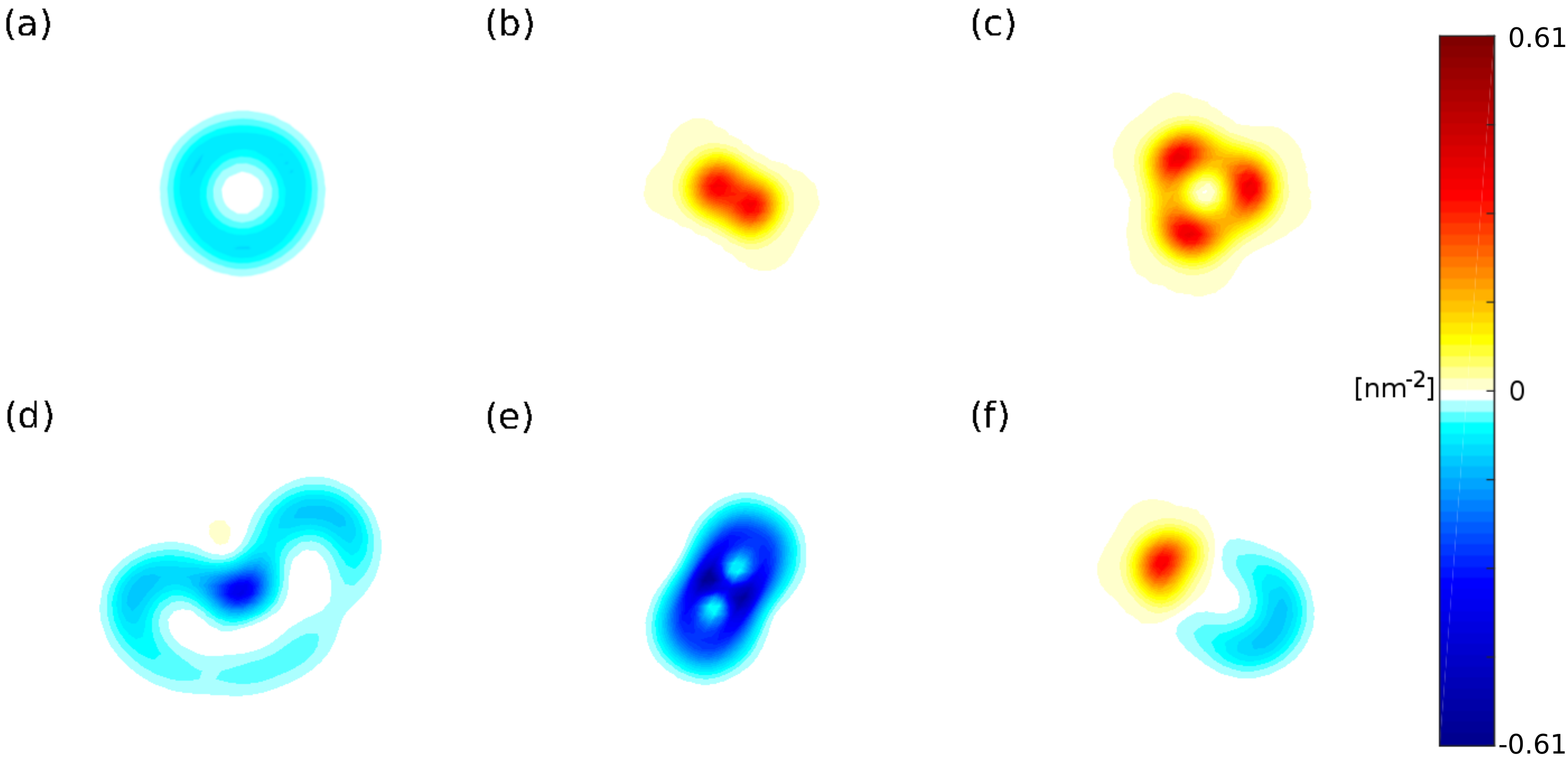}
\caption{\label{Fig4} (Color online) Topological charge densities ($\mathscr{Q}$) of the set of skyrmionic spin configurations
shown in Fig.\ \ref{Fig1}. Red and blue colors respectively denote positive and negative signs. Note that the sign of the
local vorticity ($\mathscr{M}$) is the opposite of the sign of the topological charge density in the given spin structures,
see Eq.\ (\ref{Eq_m3}).}
\end{figure}

\begin{figure*}[t]
\includegraphics[width=1.00\textwidth,angle=0]{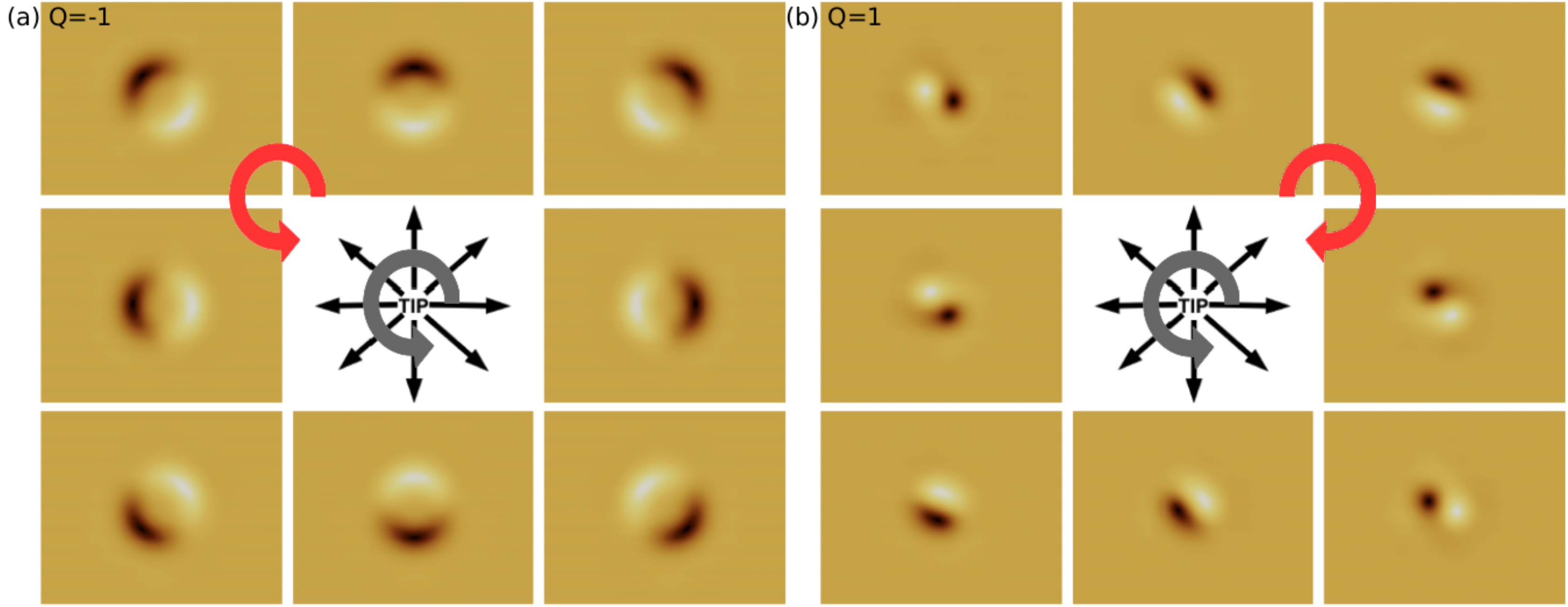}
\caption{\label{Fig5} (Color online) Calculated SP-STM images by rotating the tip magnetization direction in the surface plane
(denoted by gray circular arrows) using $P_{eff}=+0.4$ for: (a) the skyrmion with $Q=-1$; (b) the antiskyrmion with $Q=1$.
The opposite directions of contrast rotation are illustrated by red circular arrows in the two cases.}
\end{figure*}

\begin{figure*}[t]
\includegraphics[width=1.00\textwidth,angle=0]{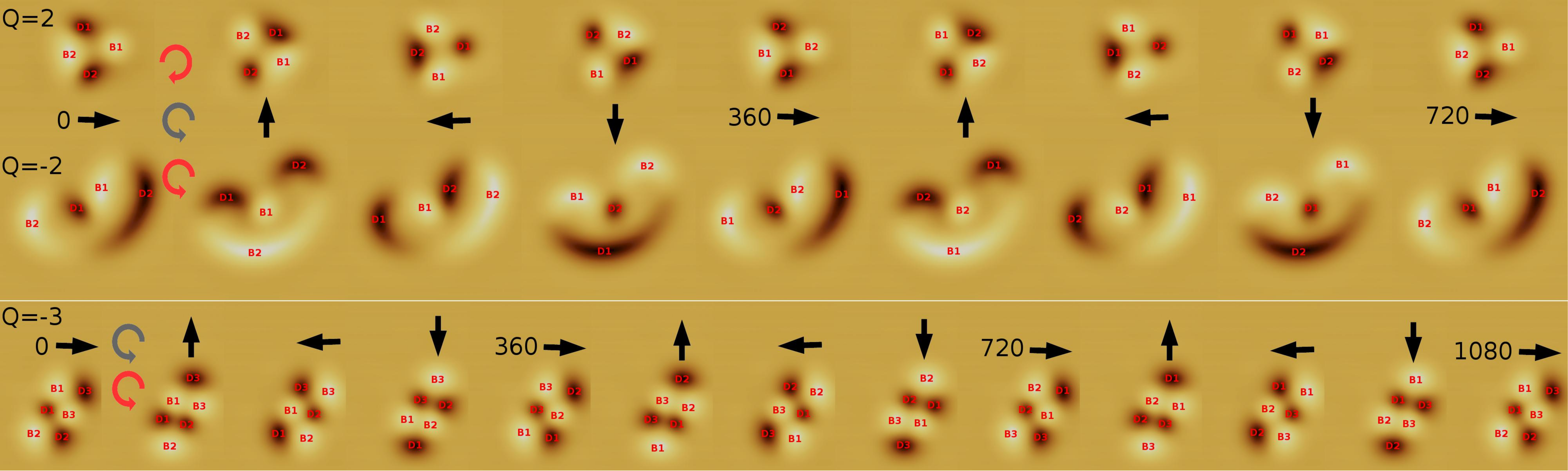}
\caption{\label{Fig6} (Color online) Calculated SP-STM images by rotating the tip magnetization direction in the surface plane
(denoted by gray circular arrows) using $P_{eff}=+0.4$ for the $Q=2$, $Q=-2$ and $Q=-3$ skyrmionic structures.
The contrast rotation is illustrated by red circular arrows in each case. The individual contrast regions are labeled
as B1-B$|Q|$ for bright and D1-D$|Q|$ for dark, and their positions upon the tip magnetization rotation are shown. Note that the
same image is obtained at a 360 degrees of tip magnetization rotation with misplaced individual contrast regions, and a $|Q|*360$
degrees tip magnetization rotation is needed to obtain the same positions of B1-B$|Q|$ and D1-D$|Q|$ contrast regions.}
\end{figure*}

\begin{figure}[t]
\includegraphics[width=1.00\columnwidth,angle=0]{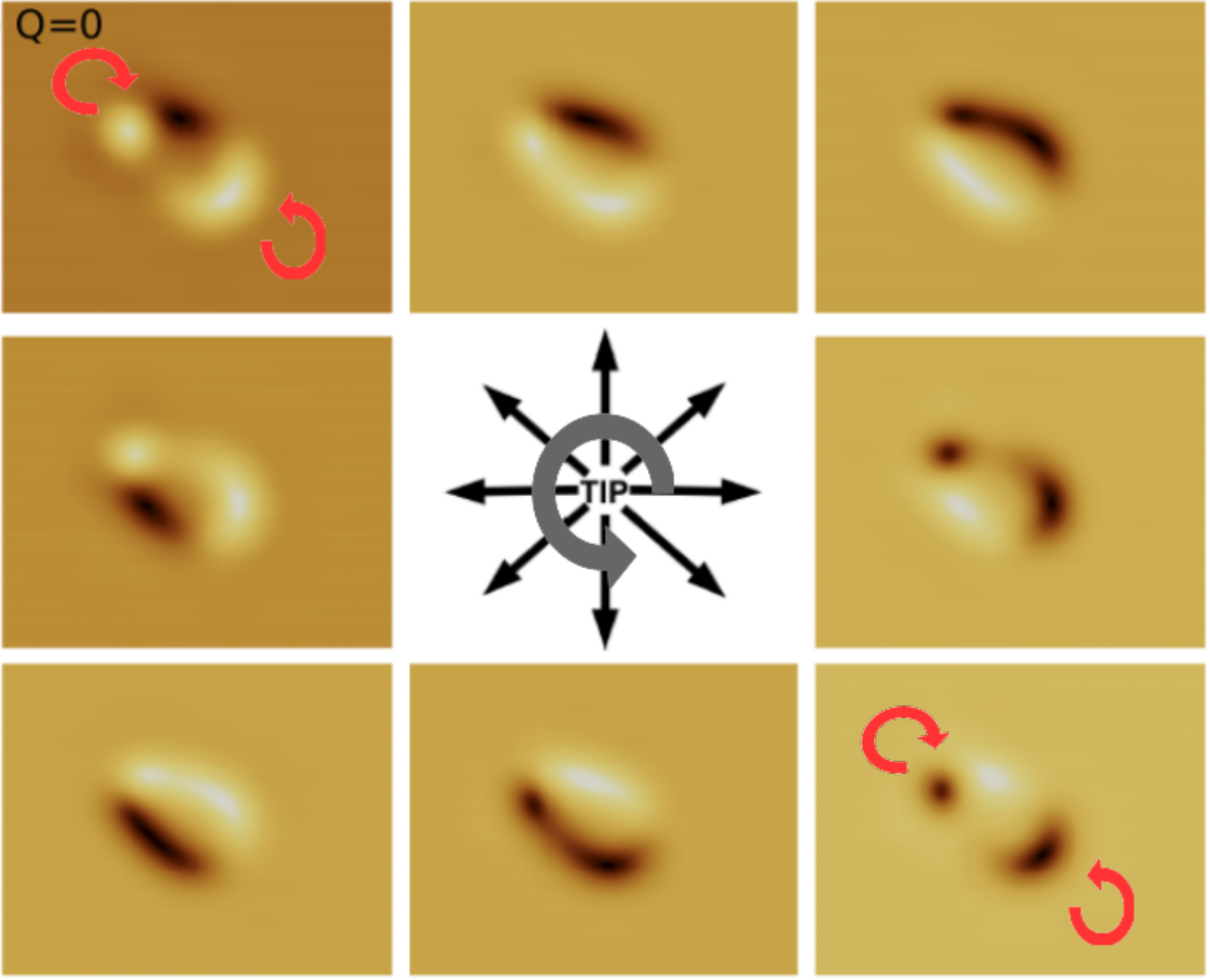}
\caption{\label{Fig7} (Color online) Calculated SP-STM images by rotating the tip magnetization direction in the surface plane
(denoted by a gray circular arrow) using $P_{eff}=+0.4$ for the ``chimera'' skyrmion with $Q=0$. The opposite
directions of contrast rotation of the skyrmionic (lower right) and antiskyrmionic (upper left) parts of the ``chimera'' skyrmion
are illustrated by red circular arrows.}
\end{figure}

\begin{figure}[t]
\includegraphics[width=1.00\columnwidth,angle=0]{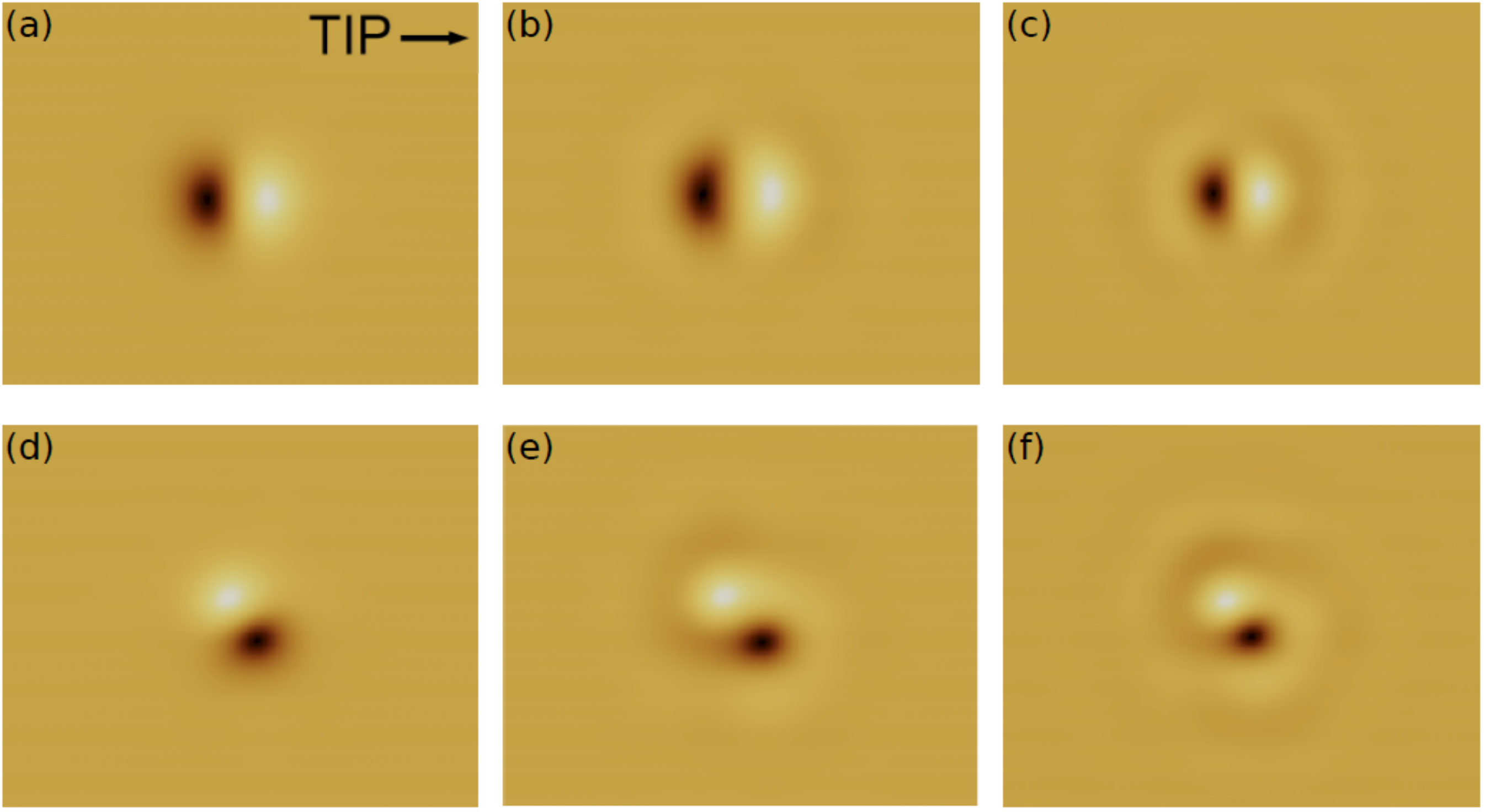}
\caption{\label{Fig8} (Color online) Simulated SP-STM images using an in-plane magnetized tip (pointing to the $+x$ $[1\bar{1}0]$
direction) with $P_{eff}=+0.4$ for the skyrmion with $Q=-1$ (a)-(c) and for the antiskyrmion with $Q=1$ (d)-(f) at different Ir
concentrations in the (Pt$_{1-x}$Ir$_{x}$)Fe/Pd(111) ultrathin magnetic film (external $B$ field values are given in parentheses):
$x=0.00$ (0.00 T) (a),(d); $x=0.10$ (4.22 T) (b),(e); $x=0.20$ (17.84 T) (c),(f).}
\end{figure}

Figure \ref{Fig1} shows a set of metastable skyrmionic spin structures obtained in a (Pt$_{0.95}$Ir$_{0.05}$)Fe/Pd(111) ultrathin
magnetic film, where attractive skyrmions have recently been reported \cite{rozsa-sk2}. Formation and stability of the metastable
structures are extensively discussed in Ref.\ \cite{rozsa-sk3}. Note that the configurations
shown in Figs.\ \ref{Fig1}(a)-(c) (first row) have also been reported by Dup\'e et al.\ \cite{dupe16stm}. In the following,
we focus on the comparison of our obtained SP-STM contrasts with the work of Dup\'e et al., and also generalize
some of their findings based on our extended set of results on higher-order skyrmions. The characterization of the skyrmionic
structures is performed by SP-STM calculations considering different fixed tip magnetization orientations, which are
sensitive to the in-line orientation of the local magnetization of the complex surface spin structures, causing the magnetic
contrast depending on the sign of $P_{eff}$. We note that the setting of an arbitrary tip magnetization orientation through a 3D
vector field is possible in SP-STM experiments according to Ref.\ \cite{meckler09}.

SP-STM images using an out-of-plane magnetized tip (pointing to the $+z$ $[111]$ direction) are shown in Figure \ref{Fig2}.
In case of a positive $P_{eff}$ the ferromagnetic background provides a bright contrast, and the skyrmionic spin structures are
imaged as dark regions exhibiting different shapes. The latter finding is in striking difference with Ref.\ \cite{dupe16stm}.
Strictly speaking, we find a circular contrast for the skyrmion with $Q=-1$ [Fig.\ \ref{Fig2}(a)] only. This circular contrast is
distorted for the antiskyrmions: $Q=1$ shows slightly elongated contrast along a specific axis [Fig.\ \ref{Fig2}(b)],
and $Q=2$ shows a rounded triangular contrast [Fig.\ \ref{Fig2}(c)]. The higher-order skyrmions with $Q=-2$ and $Q=-3$ and the
``chimera'' skyrmion with $Q=0$ show non-circular contrasts with an out-of-plane magnetized tip [Figs.\ \ref{Fig2}(d)-(f)].
These observed SP-STM contrasts are in correspondence with the symmetry (cylindrical for $Q=-1$ and $C_{|1+Q|}$ for
$Q\ne -1$), alignment and distortion of the real-space spin structures, discussed in
Ref.\ \cite{rozsa-sk3}. Note that the SP-STM contrasts reported in Fig.\ \ref{Fig2} are reversed using an out-of-plane
magnetized tip pointing to the $-z$ direction keeping the sign of $P_{eff}$, or keeping the tip magnetization direction in $+z$
and reversing the sign of $P_{eff}$.

We observe a wider variety of SP-STM contrasts when the tip magnetization is changed from out-of-plane to in-plane.
SP-STM images using an in-plane magnetized tip (pointing to the $+x$ $[1\bar{1}0]$ direction) are shown in Figure \ref{Fig3}.
The two-lobes contrast in Figs.\ \ref{Fig3}(a),(b) has been experimentally observed \cite{romming15prl} and the four-lobes
contrast in Fig.\ \ref{Fig3}(c) has been calculated \cite{dupe16stm} earlier. The other spin structures provide more complicated
contrast patterns using an in-plane magnetized tip [Figs.\ \ref{Fig3}(d)-(f)]. In general, we find the following trend: the
number of bright and dark spots is equal for each case shown in Figs.\ \ref{Fig3}(a)-(e) with an alternating order
along the perimeter of the skyrmionic structures, and the number of one type of spots (bright or dark) equals the absolute
value of the topological charge, $|Q|$.
This can be understood from the definition of the vorticity in Eq.\ (\ref{Eq_m2}), which counts how many times are
the spins rotated in the plane when passing around the perimeter of the skyrmionic structure once.
Given the fixed tip magnetization direction, this selects the number $|Q|$ of bright and dark contrast regions corresponding to
parallel and antiparallel alignments of the surface spin structure with respect to the magnetization direction of the tip
(for $P_{eff}>0$). To provide a deeper insight, Fig.\ \ref{Fig4} shows calculated topological charge densities of all
considered skyrmionic spin structures. Using the negative proportionality of the local vorticity and the topological charge
density in Eq.\ (\ref{Eq_m3}), we find uniform sign of the local vorticities in each of Figs.\ \ref{Fig4}(a)-(e). This also means
that the in-plane rotation of the spins along the perimeter of each skyrmionic structure in Figs.\ \ref{Fig1}(a)-(e)
is respectively in the same direction.

A special case is presented in Figs.\ \ref{Fig1}-\ref{Fig4}(f): the ``chimera'' skyrmion with $Q=0$.
In Fig.\ \ref{Fig3}(f) one bright and two dark spots are observed because the in-plane rotation of the spins along the
perimeter is changing direction [see the in-plane spin components in Fig.\ \ref{Fig1}(f)] providing
local vorticities of opposite sign at the skyrmionic (lower right) and antiskyrmionic (upper left) parts of the ``chimera''
skyrmion [see Fig.\ \ref{Fig4}(f)], resulting in total in zero $Q$. We return to this case after studying the effect of the
tip magnetization rotation on the in-plane SP-STM contrasts of the various skyrmionic structures.

Figures \ref{Fig5}(a) and \ref{Fig5}(b) show SP-STM images of the skyrmion in Fig.\ \ref{Fig1}(a) with $Q=-1$ and of the
antiskyrmion in Fig.\ \ref{Fig1}(b) with $Q=1$, respectively, by rotating the tip magnetization direction (black arrows) in
the surface plane indicated by gray circular arrows. The contrast rotation is illustrated by red circular arrows for the two
cases. Upon rotating the tip magnetization direction in steps of $\Delta\varphi_{TIP}=45^{\circ}$, we find that the
contrast maximum and minimum always rotate by $\Delta\varphi=45$ degrees for the skyrmion [Fig.\ \ref{Fig5}(a)].
Hence the two-lobes contrast rotates co-directionally in phase with the tip magnetization rotation for the skyrmion, in agreement
with Ref.\ \cite{dupe16stm}. According to Eq.\ (\ref{Eq_v}), this corresponds to a constant value of $\mathscr{M}=1$ along the
path of the contrast rotation.
For the antiskyrmion in Fig.\ \ref{Fig5}(b), by rotating the tip magnetization direction from $\varphi_{TIP}=0^{\circ}$ to
$360^{\circ}$ in steps of $\Delta\varphi_{TIP}=45^{\circ}$ (i.e. starting from the center right subfigure and following the
anti-clockwise rotation of the grey circular arrow), the contrast maximum and minimum rotate in order by
$\Delta\varphi=-50^{\circ},-60^{\circ},-40^{\circ},-30^{\circ},-50^{\circ},-60^{\circ},-40^{\circ}$, and $-30^{\circ}$.
This corresponds to an anti-directional two-lobes contrast rotation, in agreement with Ref.\ \cite{dupe16stm}.
According to Eq.\ (\ref{Eq_v}), $\mathscr{M}$ varies between $-1.50(=45^{\circ}/-30^{\circ})$ and $-0.75(=45^{\circ}/-60^{\circ})$
along the contrast rotation path, while an anti-phase rotation would correspond to a constant local vorticity of $\mathscr{M}=-1$.
The same findings hold when reversing the sign of $P_{eff}$ (not shown) since in that case the individual image contrasts are
inverted, and this does not affect the rotation direction of the SP-STM contrasts upon in-plane tip magnetization rotation.
This means that the SP-STM contrast rotation rule first identified by Dup\'e et al.\ \cite{dupe16stm} is insensitive to the sign
of $P_{eff}=P_S P_T$ and, thus, to the sign of the surface and tip spin polarizations, $P_S$ and $P_T$, respectively.

Let us now consider cases when $|Q|\ne 1$. Figure \ref{Fig6} shows SP-STM images of the antiskyrmion in
Fig.\ \ref{Fig1}(c) with $Q=2$, of the skyrmion in Fig.\ \ref{Fig1}(d) with $Q=-2$, and of the skyrmion in Fig.\ \ref{Fig1}(e)
with $Q=-3$, respectively, by rotating the tip magnetization direction (black arrows) in the surface plane indicated by
gray circular arrows. We find that the number of bright and dark spots ($|Q|$ each) and their alternating order along the
perimeter are preserved during the tip magnetization rotation, and this suggests that $|Q|$ can ideally be determined from
a single SP-STM measurement with a magnetic tip of any in-plane direction. Note that the size and shape of the different
contrast regions can be drastically different along the perimeter, the most pronounced case is $Q=-2$. The contrast rotation is
illustrated by red circular arrows for all three cases in Fig.\ \ref{Fig6}. Again, we find co-directional rotation of the contrast
with the tip magnetization rotation for the skyrmions ($Q<0$) and anti-directional rotation for the antiskyrmions ($Q>0$).
This, together with the insensitivity of the contrast rotations to the sign of $P_{eff}$ should enable the determination of the
sign of the vorticity $m$ and the topological charge $Q$ based on a series of SP-STM measurements performed with rotated
in-plane sensitive magnetic tips: a co-directional rotation of the contrast denotes $m>0$ (here $Q<0$), i.e., skyrmion,
and an anti-directional rotation of the contrast denotes $m<0$ (here $Q>0$), i.e., antiskyrmion.
Moreover, note that the contrast rotations in Fig.\ \ref{Fig6} are neither in phase ($\mathscr{M}=1$ along the path) nor in
anti-phase ($\mathscr{M}=-1$ along the path) with respect to the tip magnetization rotations due to $|Q|\ne 1$.
An approximately constant $\mathscr{M}=-2$ and $3$ along the path is respectively obtained for $Q=2$ [Fig.\ \ref{Fig4}(c)] and
$Q=-3$ [Fig.\ \ref{Fig4}(e)] only. We also find that the local angular velocity of the contrast rotation is indeed inversely
proportional to the local vorticity $\mathscr{M}$, see Eq.\ \ref{Eq_v}. Again, the most pronounced case is $Q=-2$, where a faster
and slower contrast rotation is respectively obtained for smaller and larger absolute values of the local vorticity, i.e., here
$\mathscr{M}$ is not constant along the path, compare the contrast features in the middle row of
Fig.\ \ref{Fig6} with Fig.\ \ref{Fig4}(d).
Importantly, if we assign unique identification labels of the bright (B1-B$|Q|$) and dark (D1-D$|Q|$) spots, a 360 degrees
in-plane rotation of the tip magnetization does not bring back the same spot to its original position, although the STM image
looks the same.
To achieve this, a $|Q|*360$ degrees tip magnetization rotation is needed. This can be clearly seen in the series of SP-STM images
for each considered skyrmionic structures in Fig.\ \ref{Fig6}: any uniquely identified spot arrives back to its original position
at a 2*360 and 3*360 degrees of tip magnetization rotation for $Q=\pm 2$ and $Q=-3$, respectively. These findings suggest that
$|Q|$ can be determined from a series of SP-STM measurements with in-plane rotated magnetic tips, where the position of
one particular contrast region is monitored. It is important to note that the above suggested procedures to obtain the sign and
magnitude of $Q$ in SP-STM experiments can only work in case of a uniform sign of the local vorticity throughout the whole
skyrmionic area.

A special case is the ``chimera'' skyrmion in Fig.\ \ref{Fig1}(f) with $Q=0$. SP-STM images of this spin structure are shown in
Figure \ref{Fig7} by rotating the tip magnetization direction (black arrows) in the surface plane indicated by
a gray circular arrow. Here, we find opposite local contrast rotations illustrated by red circular arrows for the
skyrmionic (lower right) and antiskyrmionic (upper left) parts of the ``chimera'' skyrmion, corresponding to different
signs of the local vorticity [see Fig.\ \ref{Fig4}(f)], the only case among the skyrmionic spin structures in Fig.\ \ref{Fig4}.
These local contrast rotations behave as described in Fig.\ \ref{Fig6}, i.e., the skyrmionic part shows co-directional
and the antiskyrmionic part shows anti-directional rotation of the contrast with respect to the rotation of the in-plane
tip magnetization direction, and the local angular velocity of the contrast rotation is inversely proportional to the
local vorticity, see Eq.\ \ref{Eq_v}. From this it uniquely follows that the number of bright and dark spots is not anymore
preserved upon the rotation of tip magnetization. In most of the images of Fig.\ \ref{Fig7}
there are either two bright and one dark spots, or the opposite composition: two dark and one bright. The existence of images with
an equal number of bright and dark contrast regions, e.g., one-one bright and dark in the lower left and upper right images in
Fig.\ \ref{Fig7}, makes the identification of the ``chimera'' skyrmion impossible from a single SP-STM measurement
using a tip with an arbitrary in-plane magnetization orientation. Nevertheless, as shown above, the in-plane rotation of the
tip magnetization is quite helpful in the identification of the local vorticity for the $Q=0$ ``chimera'' skyrmion and
expectedly for other skyrmionic structures with arbitrary shapes and complex domain wall structures by employing Eq.\ \ref{Eq_v}.

We have to stress that the shape of the magnetic skyrmions formed in anisotropic environments, e.g., in reconstructed films
\cite{hsu16elec,hagemeister16anis}, considerably affects the local vorticity and, consequently, the observed in-plane
SP-STM contrasts. Such an example is an axially non-symmetric spin structure derived from experimental SP-STM contrasts in
Ref.\ \cite{hsu16elec}, which shows both equal and unequal numbers of bright and dark contrast regions imaged with in-plane
magnetized tips. This already suggests the presence of a non-uniform sign of the local vorticity. A deeper theoretical
analysis puts forward that the observed spin structure is a magnetic skyrmion with $|Q|=1$ exhibiting a non-circular shape
\cite{hagemeister16anis}.
It is interesting to note the similarity of the out-of-plane SP-STM contrast of this reported skyrmion and our skyrmion with
$Q=-2$ [see Fig.\ \ref{Fig2}(d)]. The comparison of the rotated in-plane contrasts in the middle row of
Fig.\ \ref{Fig6} and those of Ref.\ \cite{hsu16elec}, however, excludes the $Q=-2$ skyrmion in the experiment.
This is in agreement with our above discussion since the proposed spin structure in Ref.\ \cite{hagemeister16anis} shows
changing directions of the in-plane rotation of the spins along the perimeter, and this clearly results in a non-uniform sign of
the local vorticity present in the system. The reason for the distorted shape of the skyrmions and their non-trivial local
vorticity in the experiment is the modified exchange interactions due to the reconstructed geometry of the magnetic film.
A similar effect has also been observed in another complex magnetic (spin spiral) state in a confined geometry \cite{palacio16}.

Finally, we present a series of SP-STM images in Figure \ref{Fig8}, where the evolution of the shape of the skyrmion with $Q=-1$
and of the antiskyrmion with $Q=1$ is shown as the function of the Ir concentration of the PtIr alloy overlayer.
The exchange parameters of Fe in the (Pt$_{1-x}$Ir$_{x}$)Fe/Pd(111) ultrathin films have been taken from Ref.\ \cite{rozsa-sk2}.
The different external $B$ field values at different Ir concentrations are given in the caption of Fig.\ \ref{Fig8}. These values
were selected to ensure similar diameters of the skyrmionic structures, and because the field-polarized state where the objects
are metastable is reached at higher field values for higher Ir concentrations \cite{rozsa-sk3}. Note that the magnitude and
direction of the $B$ field can drastically modify the size and shape of skyrmions \cite{romming15prl}. First of all, we find in
Fig.\ \ref{Fig8} that the primary two-lobes magnetic contrast of the two investigated skyrmionic structures does not change by
tuning the composition of the alloy overlayer in the studied range (0-20\% Ir in PtIr). However, for both skyrmionic structures
the appearance of an outer ring contrast is evident by increasing the Ir content of the overlayer: At $x=0.00$ Ir concentration
no ring contrast is present, where repulsive skyrmions are observed \cite{rozsa-sk2}; at $x=0.10$ a ring contrast appears,
which is even more pronounced at $x=0.20$ Ir concentration.
The outer ring contrast is a signature of oscillating in-plane components of the spins, which leads to short-range attractive
interactions between skyrmionic structures \cite{Leonov,rozsa-sk2}. This is physically governed by the frustration of
Heisenberg exchange interactions \cite{rozsa-sk3,rozsa-sk2} upon changing the composition of the alloy overlayer.
In agreement with Ref.\ \cite{Leonov}, our results highlight that not only skyrmions but also antiskyrmions can show an
attractive behavior that could be useful for technological exploitations.
We propose that attractive skyrmionic structures can be identified in SP-STM images taken with in-plane
magnetized tips by the presence of an outer ring contrast that, as we find, has a much smaller corrugation than the primary
magnetic contrast. It is expected that the corrugation of the outer ring contrast is directly related to the attractive potential
of the skyrmionic structures. It would be worthwile to study this effect both experimentally and theoretically in the future.

\section{Conclusions}
\label{sec_con}

We have investigated metastable skyrmionic spin structures with various topological charges ($Q=-3,-2,-1,0,+1,+2$, where the
vorticity is $m=-Q$) in the (Pt$_{1-x}$Ir$_{x}$)Fe/Pd(111) ultrathin magnetic film by means of spin-polarized
scanning tunneling microscopy (SP-STM) calculations. Based on the calculated SP-STM images, we conclude that an out-of-plane
magnetized tip already results in distinguished SP-STM contrasts for the different skyrmionic structures corresponding to
their symmetries. For systems exhibiting a uniform sign of the local vorticity, we have demonstrated that the magnitude of the
topological charge can be determined from the image contrast of a single SP-STM measurement using an in-plane magnetized tip:
the number of bright and dark contrast regions alternating along the perimeter of the skyrmionic spin structure equals $|Q|$ each.
For such systems we have also found that an in-plane tip magnetization rotation provides the sign of $Q$ independently of
the sign of the effective spin polarization in the tunnel junction: co-directional rotation of the contrast with respect to
the tip magnetization rotation indicates skyrmions ($m>0$, here $Q<0$) and anti-directional contrast rotation
antiskyrmions ($m<0$, here $Q>0$).
We also showed that an in-plane rotation of the tip magnetization by $|Q|*360$ degrees is needed to bring back the same
bright or dark contrast region to its original position in the SP-STM image, proving that the local angular velocity of the
contrast rotation is inversely proportional to the local vorticity. This finding could also be useful to determine $|Q|$ in case
of spin structures with a uniform sign of the local vorticity.

A special case has been identified, the ``chimera'' skyrmion with $Q=0$, where the number of bright and dark contrast regions is
not preserved upon in-plane tip magnetization rotation but its skyrmionic and antiskyrmionic parts characterized by local
vorticities of opposite sign obey the contrast rotation described above. We propose that a series of
experimentally measured in-plane SP-STM contrasts can help the topological identification of skyrmionic objects:
unequal numbers of bright and dark contrast regions in a single measurement or a non-preserved number of bright and dark contrast
regions upon in-plane tip magnetization rotation indicate a non-uniform sign of the local vorticity for the considered structure.
If neither of these apply then a uniform sign of the local vorticity can be stated and the sign and magnitude of $Q$ and $m$
can be determined as described above.
We also propose that theoretical calculations of geometrically modified exchange interactions will explain the
formation of distorted real-space magnetic textures and their local vorticity and SP-STM contrasts in the future.
Moreover, using an in-plane magnetized tip, we found that an outer ring contrast appears, indicative of both attractive skyrmions
and antiskyrmions due to the tuning of exchange interactions, in our case upon increasing the Ir content of the PtIr overlayer.

\section{Acknowledgments}

The authors thank Alexei N.\ Bogdanov for valuable discussions.
Financial support of the SASPRO Fellowship of the Slovak Academy of Sciences (project no.\ 1239/02/01), the Hungarian State
E\"otv\"os Fellowship of the Tempus Public Foundation (contract no.\ 2016-11) and the National Research, Development and
Innovation Office of Hungary under Project Nos.\ K115575 and PD120917 is gratefully acknowledged.

\end{document}